\documentclass[12pt,aps,amsmath,latexsym,amsfonts,letterpaper]{JHEP3}

\usepackage{graphicx}
\usepackage{epsfig}
\usepackage[all]{xy}
\usepackage{subfigure}

\input epsf.tex

\def\real{\mathbb R}
\def\comp{\mathbb C}
\def\integer{\mathbb Z}

\def\half{\textstyle{1\over2}}

\def\nn{\nonumber}

\newcommand{\be}{\begin{equation}}
\newcommand{\ee}{\end{equation}}
\newcommand{\bea}{\begin{eqnarray}}
\newcommand{\eea}{\end{eqnarray}}
\newcommand{\bml}{\begin{mathletters}}
\newcommand{\eml}{\end{mathletters}}

\renewcommand{\d}{\ensuremath{{\rm d}}}

\newcommand{\ba}{\begin{eqnarray}}
\newcommand{\ea}{\end{eqnarray}}

\newcommand{\lab}{\label}

\newcommand{\beq}{\begin{equation}}
\newcommand{\eeq}{\end{equation}}
\newcommand{\beqa}{\begin{eqnarray}}
\newcommand{\eeqa}{\end{eqnarray}}
\newcommand{\beqar}{\begin{eqnarray*}}
\newcommand{\eeqar}{\end{eqnarray*}}

\title{Singular manifolds, topology change and the dynamics of compactification.}
\author{Neil A. Butcher\footnote{email: ppxnb@nottingham.ac.uk}  and 
        Paul M. Saffin\footnote{email: paul.saffin@nottingham.ac.uk}\\
School of Physics and Astronomy, University Park, University of
Nottingham, Nottingham NG7 2RD, UK}

\date{\today}

\maketitle

\abstract{

We investigate the dynamics of the geometric transitions  
associated to compactified spacetimes. By including the dynamics of gravity we are able to
follow the evolution of collapsing cycles as they attempt to undergo a topology changing
transition. Rather than achieving this singular geometry we find that one of two scenarios
occur, depending on the initial conditions. Either a horizon forms, shielding a curvature
singularity, or the cycle re-expands after an initial contraction phase.
For the case where a horizon forms we identify the final state with a known analytic black-hole
solution. We also show use our results to demonstate a novel compactification mechanism, owing
to the asymptotic structure of this black-hole solution.

 }

\preprint{arXiv:yymm.nnnn [hep-th]}

\begin{document}

\section{Introduction} 
\lab{intro}

The term {\it string-theory} can be interpreted in a number of different ways, including 26-dimensional
bosonic string theory, 11-dimensional M-theory and 10-dimensional superstring theory. The common
element is that these theories exist in dimensions other four. Given that we only experience four
dimensions we are forced, if string-theory is correct, to explain why we do not observe these
extra dimensions. There are two known mechanisms to render extra dimensions unobservable, compactification \cite{Kaluza:1921tu}
and brane-worlds \cite{Rubakov:1983bb}; we shall be studying compactification, whereby the full spacetime
is constructed out of four large dimensions and some small compact manifold.
If we use the framework of superstring theory then the compact manifold ${\cal M}$ has six dimensions, and if we
require the low-energy theory to be supersymmetric this imposes certain restrictions on the type
of manifold ${\cal M}$ can be. In this paper we shall not be switching on any of the fluxes sourced by D-branes
and therefore we find that ${\cal M}$ must have special holonomy \cite{Candelas:1984yd}, implying that
${\cal M}$ is a Calabi-Yau space.
A problem now arises in the lack of a unique way to choose ${\cal M}$, 
there exist many different Calabi-Yau manifolds on which the compactification can take place. Some only 
differ by parameters, giving a continuous degeneracy, but some have completely different topologies,
leading to a discrete degeneracy. 
Calabi-Yau manifolds with the same 
topology are grouped into moduli spaces, with the spectrum of massless moduli
fields in the low-energy theory depending solely on the topology of ${\cal M}$.

This situation of having to choose one Calabi-Yau from the many was made more palatable when it was
realised that many of these Calabi-Yau manifolds were in fact connected by finite length paths in moduli space
\cite{Candelas:1988di,Candelas:1989ug,Greene:1996cy}, with the connecting geometry being singular. 
One way to picture these transitions between manifolds is to study the cycles within them, for example
it may be that certain cycles collapse to zero size on one side of the transition and expand as different
cycles on the other. During such a transition the Hodge numbers need not change, giving only a change of
intersection numbers, such as a flop transition where an $S^2$ collapses and re-expands as a different $S^2$.
Even though the Hodge numbers are unaffected this still constitutes a topology change and so is necessarily
singular from the geometrical perspective. Remarkably, string theory is able to make sense of these singular
geometries by the appearance of new light states corresponding to D-branes wrapping the collapsing cycles;
the dynamics of the low-energy theory of flop transitions has been studied in 
\cite{Brandle:2002fa,Jarv:2003qx,Jarv:2003qy}.
Other transitions may change the Hodge numbers, such as a conifold transition \cite{Candelas:1989js,Gwyn:2007qf}
where an $S^2$ collapses and reappears as an $S^3$, see also \cite{Gwyn:2007qf}.
Again, such a singular transformation can be made regular within string theory
\cite{Strominger:1995cz}, and the corresponding low-energy theory may be studied \cite{Greene:1996dh}
along with its dynamics \cite{Lukas:2004du,Palti:2005kv}.

We have just described one of the key uses of singular geometries within string-theory, namely they connect
together different moduli spaces. This was achieved by the appearance of extra light states coming from
the wrapped D-branes around the collapsed cycles. These ``extra'' states also play a crucial role in 
building low-energy models with chiral fermions and non-Abelian gauge fields, allowing
one to evade a no-go theorem \cite{Witten:1983ux} for low-energy chiral fermions coming from compactification
\cite{Acharya:2000gb,Atiyah:2000zz,Atiyah:2001qf,Acharya:2001gy}.

The types of singularity that have proven useful in the above constructions are conical, taking the local description
of a discrete quotient of a smooth manifold $\tilde{X}$,
$X=\tilde{X}/\Gamma$, where $\Gamma$ is a finite symmetry group; the singularity is then the
fixed point set of $\Gamma$. Of particular interest are those singularities which allow a smooth resolution, by
blowing up certain cycles of zero size contained within the fixed point set of $\Gamma$. This is achieved in practise
by a surgery which replaces a ball around the conical singularity with a 
ball of a smooth special-holonomy space \cite{Gibbons:1979xn,Page:1979zu,Joyce:2000}.
The moduli of this special-holonomy space then appear in the low-energy theory as moduli fields \cite{Lukas:2003dn}.

In this paper we study the dynamics of the resolved spaces while the cycles are collapsing. Unlike previous studies on
the dynamics of such spaces \cite{Brandle:2002fa,Jarv:2003qx,Jarv:2003qy,Lukas:2004du,Palti:2005kv} we are particularly
interested in the gravitational properties of collapsing cycles; more specifically, whether a horizon forms as the cycle
becomes small. Should a horizon form in the higher dimensional theory, then this would render the low-energy theory
based on the moduli fields inapplicable near the conical singularity, implying that the dynamics of topology changing
processes is more complicated than simply studying the dynamics of the moduli fields in the low-energy description.

To have a specific model we shall be studying the resolution of the $\comp^2/\integer_2$ orbifold. Singularities of this type
are common in the construction of compact manifolds of special holonomy \cite{Joyce:2000}, where one typically starts
with a torus, imposes some reflection symmetries, and then replaces the fixed points of the symmetries smooth manifolds.
In our case the singularity is resolved with the Eguch-Hanson instanton \cite{Eguchi:1978xp,Eguchi:1980jx}.
We begin our study in section \ref{sec:instantons} with a discussion of four-dimensional Euclidean instantons, and
introduce the Eguchi-Hanson solution along with its size modulus for the collapsing $S^2$. In section \ref{NumSim}
we describe the method we use to simulate Einstein's equations, with the initial data presented in section \ref{sec:init}.
We finish by giving our results in section \ref{results} and conclusions at the end.

\section{Gravitational instantons}
\lab{sec:instantons}

To have an explicit construction of a compact manifold with special holonomy using the methods of
Joyce \cite{Joyce:2000} one needs to have explicit Ricci-flat metrics of special holonomy; a number
of such metrics are known for non-compact manifolds \cite{Eguchi:1978xp,Gibbons:1979zt,Gibbons:1989er,Cvetic:2001zx}.
One then replaces a neighbourhood of the singularity with a region of the non-compact space, using some
form of smoothing at the join, to construct the metric on the smooth compact manifold.
As we are interested only in the dynamics near the putative conical singularity we only sudy the dynamics
of the non-compact space, assuming that information far from the region of interest has no impact
over the timescales of our simulations.

The singularity we are interested in resolving is the simplest case, $\comp^2/\integer_2$, and for that
we need to know something about Ricci-flat metrics on four-dimensional Riemannian manifolds, also
known as gravitational instantons. 

\subsection{Eguchi Hanson metric}
\lab{EHmetric}

The Eguchi-Hanson metric is a regular self-dual, hyper-K\"ahler metric in four-dimensions and has the asymptotic
structure of $\comp^2/\integer_2$ \cite{Eguchi:1978xp,Eguchi:1980jx}, 
i.e. it is a resolution of the $\comp^2/\integer_2$ conical singularity.
It is constructed as a cohomogeneity-one metric with squashed three-spheres as the level surfaces and has
the explicit form
\bea
\lab{eq:EHMet}
\d s^2_{EH}(l) &=&\alpha(\rho)^{-1}\d\rho^2 +
             \frac{1}{4} \rho^2 \left[(\sigma_1^2+\sigma_2^2) + \alpha(\rho)\sigma_3^2\right],\\
\alpha(\rho)&=&1-\left(\frac{l}{\rho}\right)^4.
\eea
We have used the conventional left-invariant one-forms of SU(2) which satisfy
\bea
\d\sigma_i&=&-\half\epsilon_{ijk}\sigma_j\wedge\sigma_k,
\eea
and the parameter $l$ is a constant parameter i.e. a modulus of the solution.

From the above form of the metric we see that there is an apparent singularity at $\rho=l$, we get a clearer
understanding of its nature if we look close to this region using the following co-ordinates,
\be
\lab{eq:Rdef}
\rho=l+\frac{R^2}{l}.
\ee
This results in the metric taking the form
\be
\d s^2 =\left[\d R^2+ R^2\sigma_3^2\right]+\frac{l^2}{4} (\sigma_1^2+\sigma_2^2),
\ee
which clearly shows that the apperent singularity at R=0 ($\rho=l$) is just a co-ordinate artefact,
and that the manifold looks locally like a product of flat space and a two-sphere of radius $l/2$;
this type of removable singularity is termed a bolt singularity \cite{Gibbons:1979xm}.
It is the finite size of this two-sphere which has resolved the singularity, by taking $l$ to zero
in (\ref{eq:EHMet}) we explictly see the metric becomes $\comp^2/\integer_2$. (The $\integer_2$ comes
from an identification required to make the origin of the resolved space regular \cite{Eguchi:1980jx}.)

\subsection{Moduli evolution.}
\lab{sec:moduli}

In our simulations we shall be working in five dimensions, with the four spatial dimensions 
initially taking the form of the Eguchi-Hanson metric. This is an exact solution of Einstein's
equations, so to get the spacetime to evolve we must give the metric some momentum. Before presenting
the full numerical approach we should see what we can find analytically.
As pointed out by Manton in the context of BPS monopoles \cite{Manton:1981mp} one can understand the
low-energy dynamics of a system by allowing its moduli to have a small time dependence. 
We introduce a time-dependent modulus $L(t)$, such that $L(t=0)=l$. In our case we can use the Einstein-Hilbert action 
\be
S=\int \d t\;\d ^4x\,\sqrt{-g} R,
\ee
with the metric
\ba
\lab{eq:metricInitial}
\d s^2&=&-\d t^2+\d s^2_{EH}\left(L(t)\right),
\ea
to derive the effective action for the modulus $L(t)$,
\ba
S_{eff}&=&\frac{\pi^2}{8}\int \d t\;\left(\frac{\d}{\d t}L^2\right)^2.
\ea
This leads to the conclusion that the moduli space approximation predicts that $L(t)^2$ evolves
linearly with time. Also, given that the boundary of the moduli space is $L=0$ we see that
the modulus $L$ can reach the boundary in finite time.

\section{Numerical Simulation}\lab{NumSim}

The idea now is to start with a regular five-dimensional metric and evolve it numerically  using 
Einstein's equations. A similar project was undertaken by Bizon {\it el al} \cite{Bizon:2005cp}
who realised that Birkhoff's
theorem could be evaded in five spacetime dimensions. Their initial metric however contained a nut singularity
at the origin as opposed to the bolt in (\ref{eq:metricInitial}). While both nut and bolt singularities are
removable co-ordinate singularities and so are regular analytically, we found that the numerical techniques of
\cite{Bizon:2005cp} broke down at the origin. After numerous attempts, using different forms for the metric, we settled
on the ADM formalism for evolving the metric \cite{Arnowitt:1962hi,Kelly:2001kj}. 
This involves three metric functions along with their momenta,
\be
\d s^2 = -\d t^2+\tilde a(t,r)^2\;\d r^2+\tilde b(t,r)^2(\sigma_1^2+\sigma_2^2)+\tilde c(t,r)^2\sigma_3^2.
\lab{eq:ADMmetric}
\ee
The ADM equations then give first order evolution equations for the metric functions and their momenta,
along with a Hamiltonian constraint and momentum constraints. Note that 
we have chosen the gauge such that the lapse is unity and shift vector vanishes, we have also enforced the
bi-axial form of the Eguchi-Hanson metric for simplicity.
It is certainly possible for (\ref{eq:ADMmetric}) to describe the metric we are interested in, (\ref{eq:metricInitial}),
however if we were to simply equate the $r$ co-ordinate of (\ref{eq:ADMmetric}) with the $\rho$ co-ordinate
of (\ref{eq:metricInitial}) we would find that $\tilde a(t,r)$ diverges at the origin. This clearly causes unacceptable
numerical problems, so we need to choose our parametrization more carefully, as desribed in the next section.

\subsection{Metric}

To aid the stability of the algorithm, particularly in establishing sensible boundary conditions, we require
that

\begin{enumerate}
\item
All three variables and all three momenta to remain even and finite at the origin.
\item
All three variables and all three momenta to tend to finite (maybe zero) values asymptotically.
\item
We minimize the amount of division by variables in all equations of motion.
\end{enumerate}

To that end we evolved the following form for the metric
\be
\d s^2 = -\d t^2 +\left[1+4\left(\frac{r}{l}\right)^2\right]e^{2A}\d r ^2 +l^2\,\left[1+\left(\frac{r}{l}\right)^4\right]e^{2B}(\sigma_1^2+\sigma_2^2) + r^2\left[1+\left(\frac{r}{l}\right)^2\right]e^{2C}\sigma_3^2
\lab{metric}
\ee

In order to impose all the necessary boundary conditions at once, keeping the equations regular at the origin,
we used techniques outlined in \cite{Alcubierre:2004gn, Ruiz:2007rs}. This involved the introduction of three new variables 
$D_A$,$D_B$ and $D_C$, to replace the spatial derivatives,
\bea
D_A&=&A'+\frac{4r}{l^2+4r^2},\\
D_B&=&B'+\frac{2r^3}{l^4+r^4},\\
D_C&=&C'+\frac{r}{l^2+r^2}.
\eea
We also introduced the momenta $K_A$,$K_B$ and $K_C$, defined as
\bea
K_A&=&-\dot{A},\\
K_B&=&-\dot{B},\\
K_C&=&-\dot{C},
\eea
where $\dot{ }$ indicates derivative with respect to time and ' indicates derivative with respect to r.
The $D_i$ were chosen so as to be odd at the origin as these were simple boundary conditions to impose.
Actually, the full set of boundary conditions at the origin may be found  by requiring local flatness
\cite{Alcubierre:2004gn}, in which case we find
\bea
A(t,r)  &\sim& A^0(t)  +\mathcal{O} (r^2),\\
D_A(t,r)&\sim&          \mathcal{O} (r),  \\
K_A(t,r)&\sim& K_A^0(t)+\mathcal{O} (r^2),
\eea
with similar relations for the functions $B(t,r)$, $C(t,r)$, $K_B(t,r)$, $K_C(t,r)$ at the origin. We also find that
\be
A^0(t)=C^0(t),\quad K_A^0=K_C^0.
\ee
By giving the metric functions some initial momentum
the spatial part of the metric will cease to remain Eguchi-Hanson, however it 
will retain some Eguchi-Hanson features, 
at least for early times. Notably, the bolt singularity at the origin will remain, still describing a two-sphere
of radius $L(t)/2$.
We used the value of B at the origin to define this $L(t)$ at later times.
\be
L(t)=2\exp(B^0(t))
\ee
Note that for $L(t)$ to vanish, then $B(t,r=0)$ must diverge.

\section{Initial conditions}
\lab{sec:init}

In the parametrization of (\ref{metric}), using the co-ordinate $r=R$ of (\ref{eq:Rdef})(\ref{eq:EHMet})
we find that our initial conditions for the metric functions take the form
\bea
\lab{eq:ABC}
A&=&\frac{1}{2}\,ln\left[4\,\left(\frac{r}{l}\right)^2\,\left(1-\left(\frac{l^2}{l^2+r^2}\right)^4\right)^{-1}\,\left(1+4\,\left(\frac{r}{l}\right)^2\right)^{-1}\right]\nn\\
B&=&\frac{1}{2}\,ln\left[\frac{1}{4}\,\left(1+\left(\frac{r}{l}\right)^2\right)^2\,\left(1+\left(\frac{r}{l}\right)^4\right)^{-1}\right]\nn\\
C&=&\frac{1}{2}\,ln\left[\frac{1}{4r^2}\,\left(1-\left(\frac{l^2}{l^2+r^2}\right)^4\right)\,\left(l+\frac{r^2}{l}\right)^2\,\left(1+\left(\frac{r}{l}\right)^2\right)^{-1}\right].\nn\\
\eea
If we impose vanishing momenta then this would constitute an exact solution of the equations of motion.
As a check of our numerics we do indeed find that the system remains static. As we want to evolve the
Eguchi-Hanson metric toward the conical singularity we must impose some non-vanishing momentum for the
metric functions. This is not a completely trivial task given that general relativity imposes constraints
coming from the gauge fixing (appendix \ref{app:constraints}). The two constraints, Hamiltonian and momentum, mean that
once $A(t=0,r)$, $B(t=0,r)$ and $C(t=0,r)$ are fixed according to (\ref{eq:ABC}) there is one free function left
to describe the momentum. To fix this function we take our motivation from the moduli space approximation
of section \ref{sec:moduli} and find that initially we have
\be
K_B=-\dot{B}=-\frac{\dot{L}}{l}\,\left(\frac{l^2-r^2}{l^2+r^2}\right),
\ee
so we are able to choose an $\dot L$ and derive from this $K_B$. we imposed that $\dot L$ was required to be:
\begin{enumerate}
\item
even at the origin.
\item
finite and negative at the origin.
\item
exactly zero far from the origin.
\item
continuous and differentiable to first order.
\end{enumerate}
The first condition ensures that $K_B$ is even, the second means that we push the Eguchi-Hanson space towards
the conical singularity. The third condition is imposed so that only the form near the origin is important,
and that the non-compact nature of Eguch-Hanson does not affect the evolution. The final condition gives a smooth
profile for us to evolve.

Only one of the three momenta, $K_B$, was specified explicitly by $\dot{L}$, with the 
other two being derived from the constraints $(\ref{HamCon},\ref{MomCon})$ using a 4th order Runge-Kutta algorithm.

We decided on an $\dot{L}$, taking the form:
\bea
\lab{eq:Ldot}
\dot L&=&
\left\{
\begin{array}{cc}
-\dot{L_0}\,(1-(\frac{r}{r0})^2)^2\quad & r<r0 \\
0 & r>r0
\end{array}
\right.
\eea

where $\dot{L_0}$ is a positive constant (the magnitude of $\dot{L}$ at the origin) and r0 is another constant 
which determines the outer radius of the non-zero $\dot{L}$. With the initial data fixed we now evolve the system
according to the equations of motion laid out in appendix \ref{app:eom}. To acheive this we used a 4th order Runge-Kutta algorithm.
while keeping a check that the constraints of appendix \ref{app:constraints} remained small; typically they were of order 0.005.

\section{Results}
\lab{results}

\subsection{Apparent horizons}
The outgoing radial null geodesics can usually be seen to be increasing in area, but after the formation of the 
apparent horizon, which is a null trapped surface, the null geodesics are no longer increasing in area all null 
rays are in fact converging.
This apparent horizon shows that there exist points in space for which all null geodesics are unable to diverge 
to $\mathcal{I}^{+}$. These points are therefore behind an event horizon. This event horizon must include everything 
within the apparent horizon (and maybe more), however the apparent horizon is far easier to detect and it proves the 
existence of the event horizon $\cite{Thornburg:1995cp}$.
In order to determine whether a black hole has formed we continually looked for such an apparent horizon.

At the apparent horizon, a congruence of null geodesics no longer increase their area and we may write,
\bea
0&=&\left[\frac{d\,Area}{dt}\right]_{null},\\
&=&\left[\frac{\partial \,Area}{\partial t}\right]_{r}
  +\left[\frac{\partial \,Area}{\partial r}\right]_{t}\,\left[\frac{{\rm d}r}{{\rm d}t}\right]_{null}.
\eea
The area is determined by $Area\sim\tilde{b}^2\tilde{c}$ in terms of (\ref{eq:ADMmetric}), with 
$\left[\frac{{\rm d}r}{{\rm d}t}\right]_{null}=1/\tilde{a}$. This may also be calculated in terms of the
extrinsic curvatures $\cite{Thornburg:1995cp}$ to give
\bea
0&=&-2\,K_B-K_C+\frac{2\,D_B+D_C+\frac{1}{r}}{\sqrt{1+4\left(\frac{r}{l}\right)^2}\,\,e^{A}},
\eea
at an apparent horizon.
\subsection{Formation of black holes}

Depending upon our input parameters, $\dot{L}$ and r0, there were three possible outcomes to adding the momentum:
\begin{enumerate}
\item
For sufficiently low $\dot{L}$ and r0, there was insufficient initial momentum to observe the creation of either a black hole or a singular topology.
\item
For an intermediate range in the parameters the system produced an apparent horizon. After initially increasing,
the area of the apparent horizon converged to a constant value.
\item
For large initial parameters the system already contained an apparent horizon simply due to the initial conditions.
\end{enumerate}

Our interest is directed to case 2. In this case a black hole forms, which was not initially present.

Before we present the results for a range of parameters we focus on a single example where we took $r0=1.0$, $\dot{L_0}=2.3$.
In Fig. \ref{fig:L}a we plot the time dependence of $L$ as measured at the origin, and also give the area of
the horizon - both in units of $l$. What the figure shows is that $L$ is monotonically decreasing, with $L^2$
decreasing approximately linearly in the initial phase. This is consistent with the expectations from the moduli
space approximation of section \ref{sec:moduli}. However, at some point an apparent horizon forms ($t\sim 0.11$ in
the simulation) which then implies an event horizon exists - something the moduli space approximation does not
account for. At this point we can no longer trust any low-energy dynamics derived from the moduli space approximation.
We also see from Fig. \ref{fig:L}a that the area of the apparent horizon increases initially, but the settles down
to a fixed value. Presumably this corresponds to the formation of what would become a static black hole; we shall
discuss this further in section \ref{sec:blackhole}. The figure also shows that $L$, as defined at the origin, reaches
zero in finite co-ordinate time $t$ ($t\sim 0.3$ in
the simulation). This corresponds to a divergence in the metric function $B(t,r)$ and is in
fact a curvature singularity. Fortunately this is hidden behind the horizon.

In order to get a clearer understanding of the causal structure
of our solution we present in Fig. \ref{fig:L}b a plot showing various radial outgoing null geodesics. Superimposed
on this is the curve showing the location of the apparent horizon. We see that initially the null rays continue outwards
and, given the asymptotically locally flat structure of Eguchi-Hanson, reach null infinity. However, some time later
the outgoing null rays near the origin turn around and head towards $r=0$. The presence of such null rays indicates
that a horizon has formed, and this is confirmed by the existence of the apparent horizon.

\FIGURE{
\includegraphics[width=8cm, height=8cm]{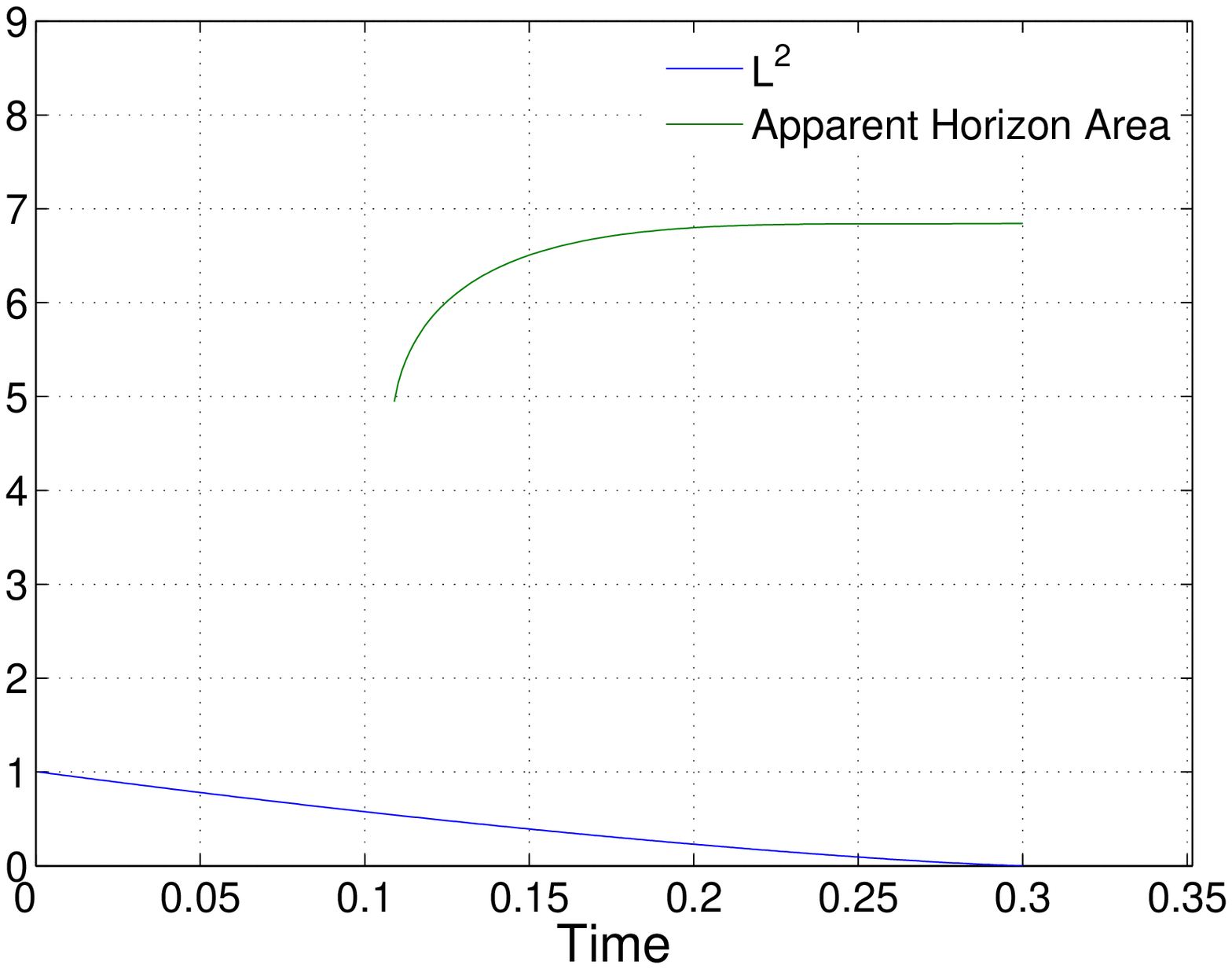}\includegraphics[width=8cm, height=8cm]{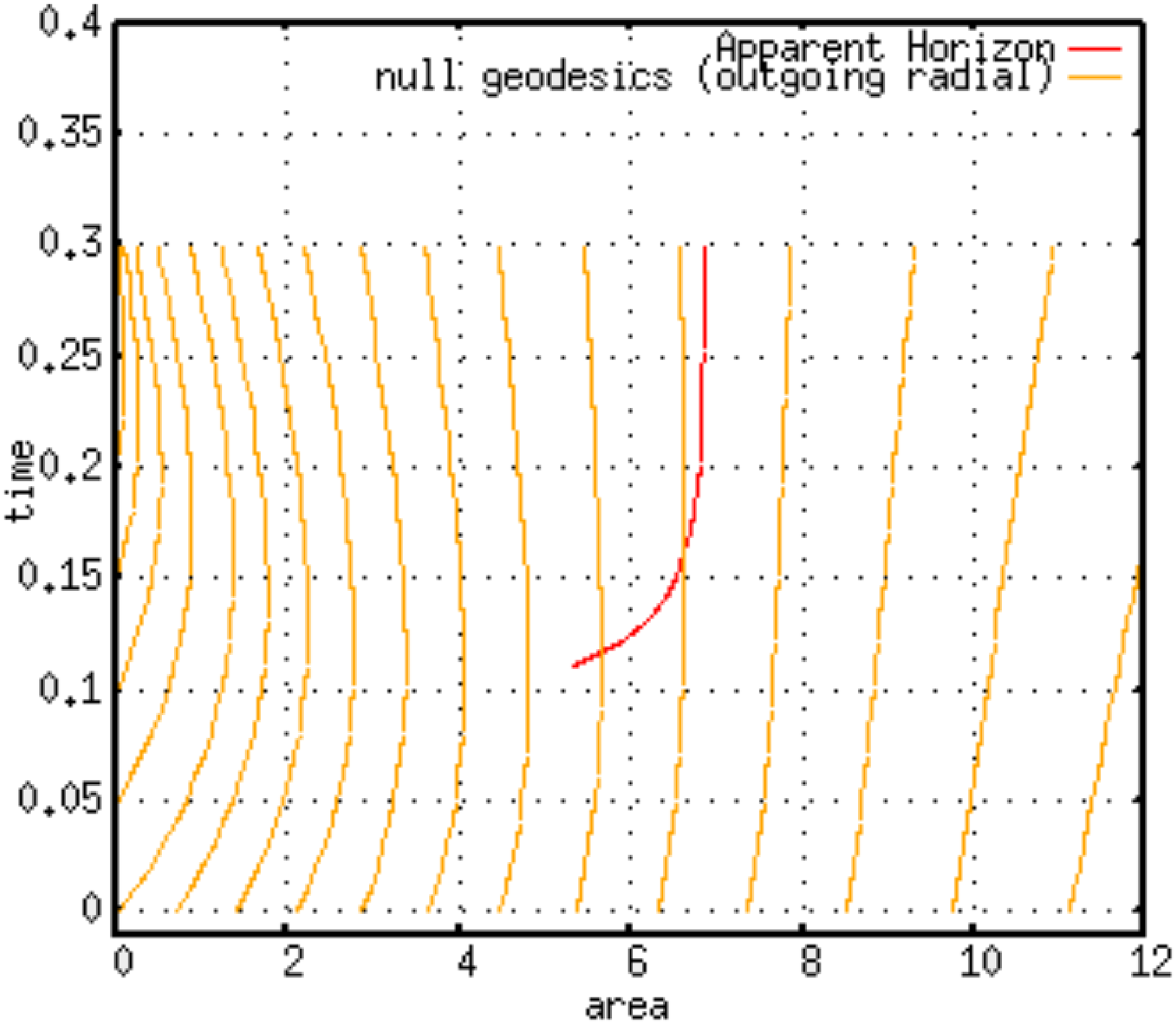}
\caption{(a) The steady decrease of $L^2$ and the formation of the apparent horizon.
         (b) The outgoing radial null geodesics and the apparent horizon which later formed.} 
\lab{fig:L}
}



\subsection{The resulting black hole}
\lab{sec:blackhole}

By the time the program ends (due to the curvature singularity at $r=0$)
the apparent horizon has settled to a single area which can be measured. The natural question is
``what is the final state?'' Given that we cannot run the simulations beyond the curvature
singularity we can only offer a conjecture to answer this question. However, given that the horizon has
converged to a constant value we believe that it is reasonable to suggest a five-dimensional black-hole
is in the process of forming. The black-hole which fits our requirements was written down in its
Kaluza-Klein dimensionally reduced form in \cite{Gibbons:1985ac,Dobiasch:1981vh}.
Written in its five-dimensional form this black hole looks like \cite{Ishihara:2005dp}
\bea
\d s^2&=&-f\d t^2+\frac{k^2}{f}\d r^2+\frac{r^2}{4}\left[k(\sigma_1^2+\sigma_2^2)+\sigma_3^2\right],\\
f(r)&=&\frac{(r^2-r_+^2)}{r^2}\\
\lab{eq:KKlein}
k(r)&=&\frac{(r_\infty^2-r_+^2)r^2_\infty}{(r_\infty^2-r^2)^2}
\eea
and describes a static black-hole with a squashed three-sphere for a horizon
at $r=r_+$. The radial co-ordinate range
is $0<r<r_\infty$ and the parameter range is $0<r_+<r_\infty$. If we accept that this is the end state of the 
Eguchi-Hanson collapse then we are free to evaluate the squashing function $k(r)$ at the horizon, provided it 
too has settled to a single value before the program's end.

The asymptotic structure of the black-hole is interesting in that it is not asymptotically flat, rather it is
asymptotically locally flat and takes the form \cite{Ishihara:2005dp}
\bea
\d s^2&=&-\d T^2+\d R^2+R^2\d \Omega_{S^2}+\frac{r_\infty^2}{4}\chi^2.
\eea
So, locally this looks like $\real^{(1,3)}\times S^1$, where the circle has radius $r_\infty/2$. we can find the parameters
$r_+$ and $r_\infty$ by evaluating the area of the horizon, and the squashing parameter on the horizon($k(r_+)=k_+$),
\bea
r_+&\sim&\left(area/k_+^2\right)^\frac{1}{3}\\
r_\infty&=&r_+\sqrt{\frac{k_+}{k_+-1}}
\eea

This result gives us a rather novel method for dynamical compactification. Suppose that instead of starting with 
a compact manifold, where a portion of Eguchi-Hanson space has been glued in, we start with the full Eguchi-Hanson
space with its four ``large'' spatial dimensions. Then our results show that this evolves to a space where one of
the spatial dimensions compactifies to a circle, giving three ``large'' dimensions and one ``small''.

\subsection{Varying the initial data.}

The black hole's area and the extent to which the angular part is squashed depends on the initial conditions;
in our parametrization (\ref{eq:Ldot}) this means changing $\dot{L_0}$ and $r0$.
The results of varying $\dot{L_0}$ while keeping $r0$ constant are shown in Fig. \ref{fig:area}.

As described earlier, for values of $\dot{L}$ which are too small the system never produces any sort of horizon, no singularity 
is formed and the value of L drops for a small amount then begins to rise again, there is not enough energy to form a black hole 
or a singularity.

Alternatively, if we take $\dot{L}$ to be too large then the initial data already contains an apparent horizon, rather
than forming one dynamically. The results we present in Fig. \ref{fig:area} cover the intermediate range where there
is enough localized energy to form a black hole, but not so much that it is there at the start of the simulation.

In Fig. \ref{fig:area}, where $r0=1.0$, the apparent horizon forms dynamically for $0.7<\dot{L_0}<2.8$ and it's area can be 
seen to converge and be measured. 
Over the duration of the simulation
the squashing parameter was seen to converge for the range 
$2.0<\dot{L_0}<2.8$, and in all these cases 
it converges to a value greater than one. 
This is consistent with the numerical squashing parameter being identified with the analytic form of $k_+$, given 
in (\ref{eq:KKlein}), which must also remain greater than one at the horizon.

\FIGURE{\includegraphics[width=8cm, height=8cm]{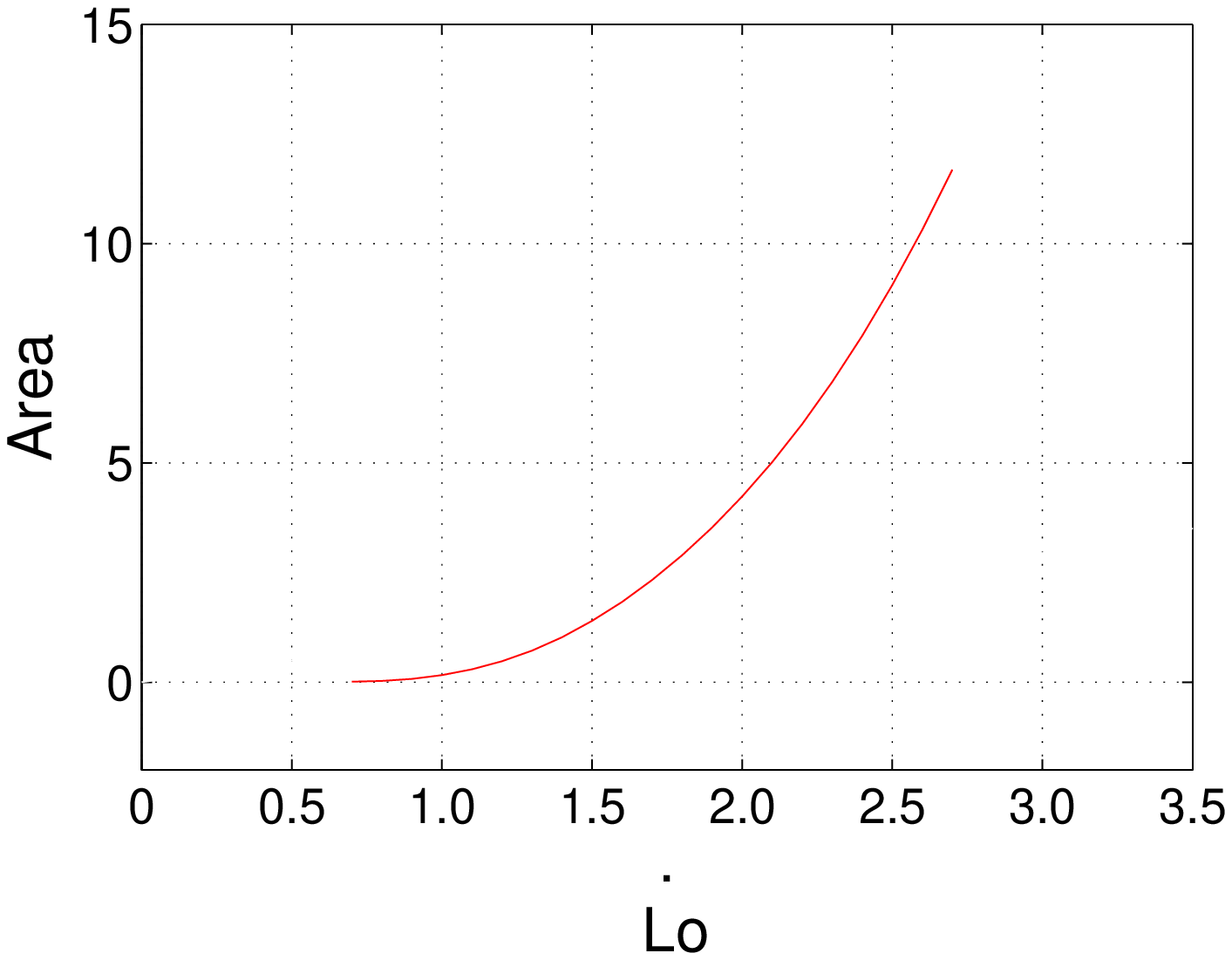}\includegraphics[width=8cm, height=8cm]{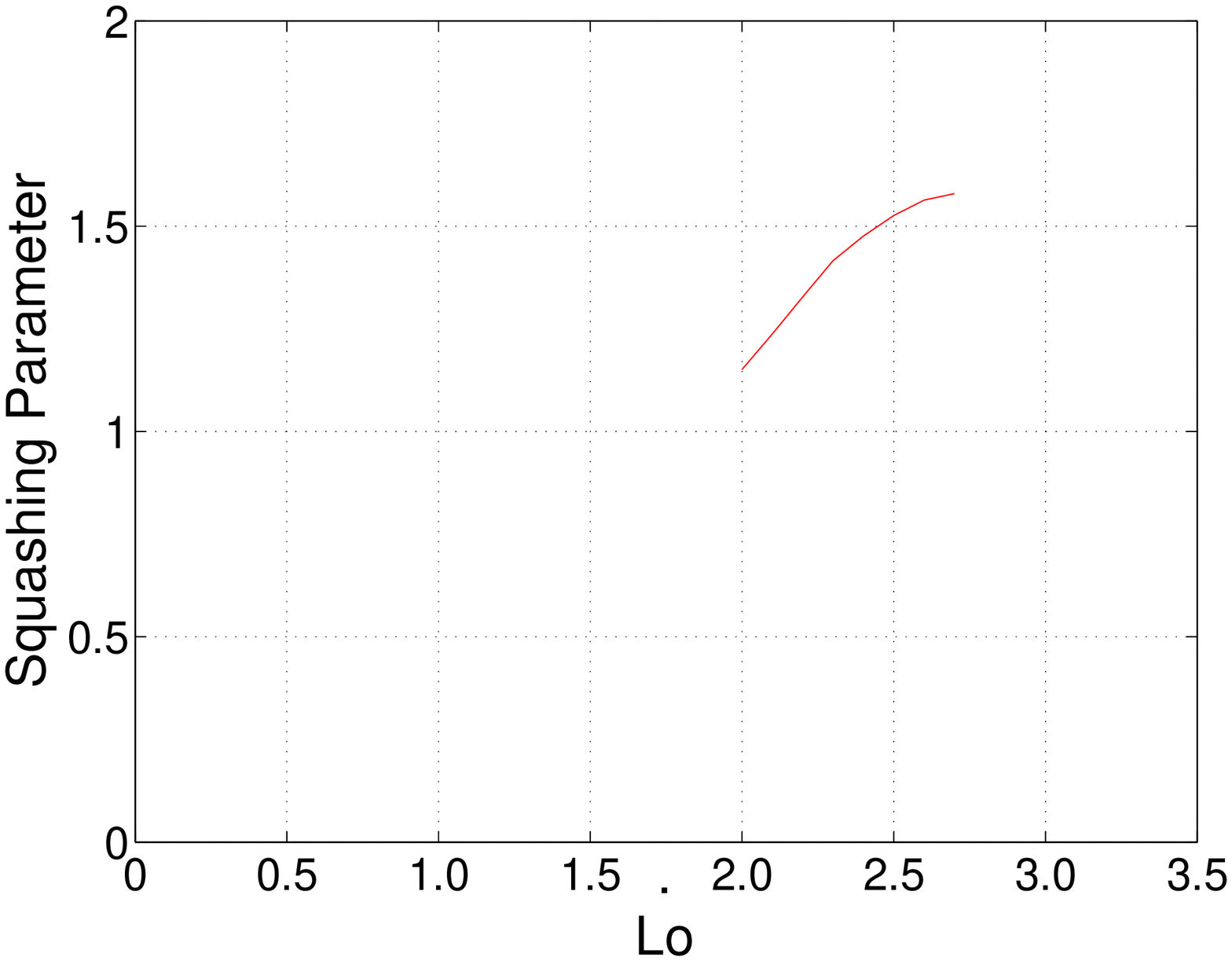}
\caption{The effect on the final area and extent of squashing (only for values for which it has converged) at the horizon
         due to differing the initial $\dot{L_0}$ ($r0=1.0$).} \lab{fig:area}}

If we vary both the values of $\dot{L_0}$ and $r0$ we can find the resulting area in a great many cases. this is shown in Fig.
\ref{fig:param}. The range in which a horizon forms at a late time is shown and given a shading scale to indicate the area 
of that horizon. Within the region marked A, there is insufficient energy to form a horizon at all. Region B marks the existence 
of a horizon within the initial conditions. 

\FIGURE{\includegraphics[width=8cm, height=8cm]{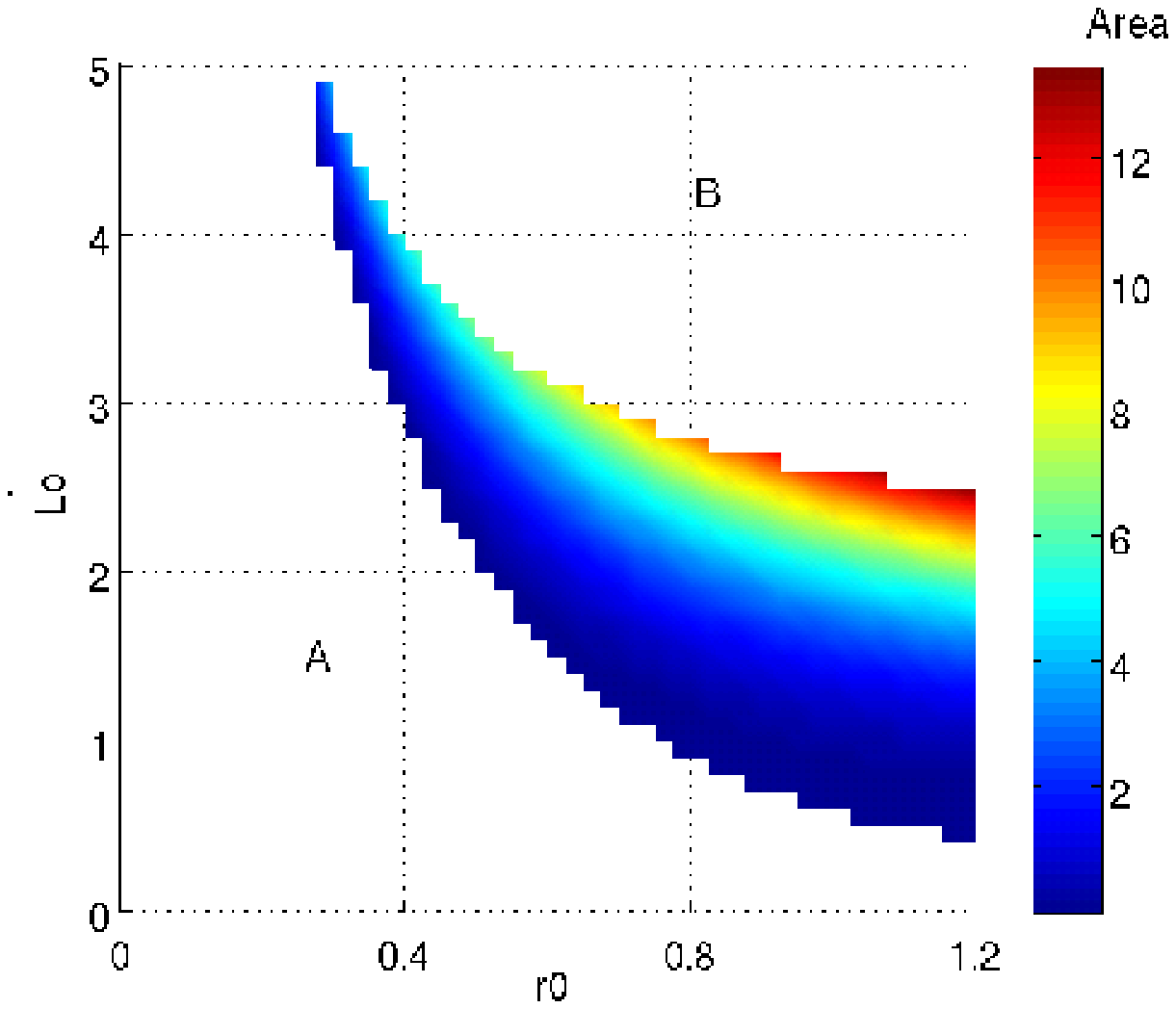}\includegraphics[width=8cm, height=8cm]{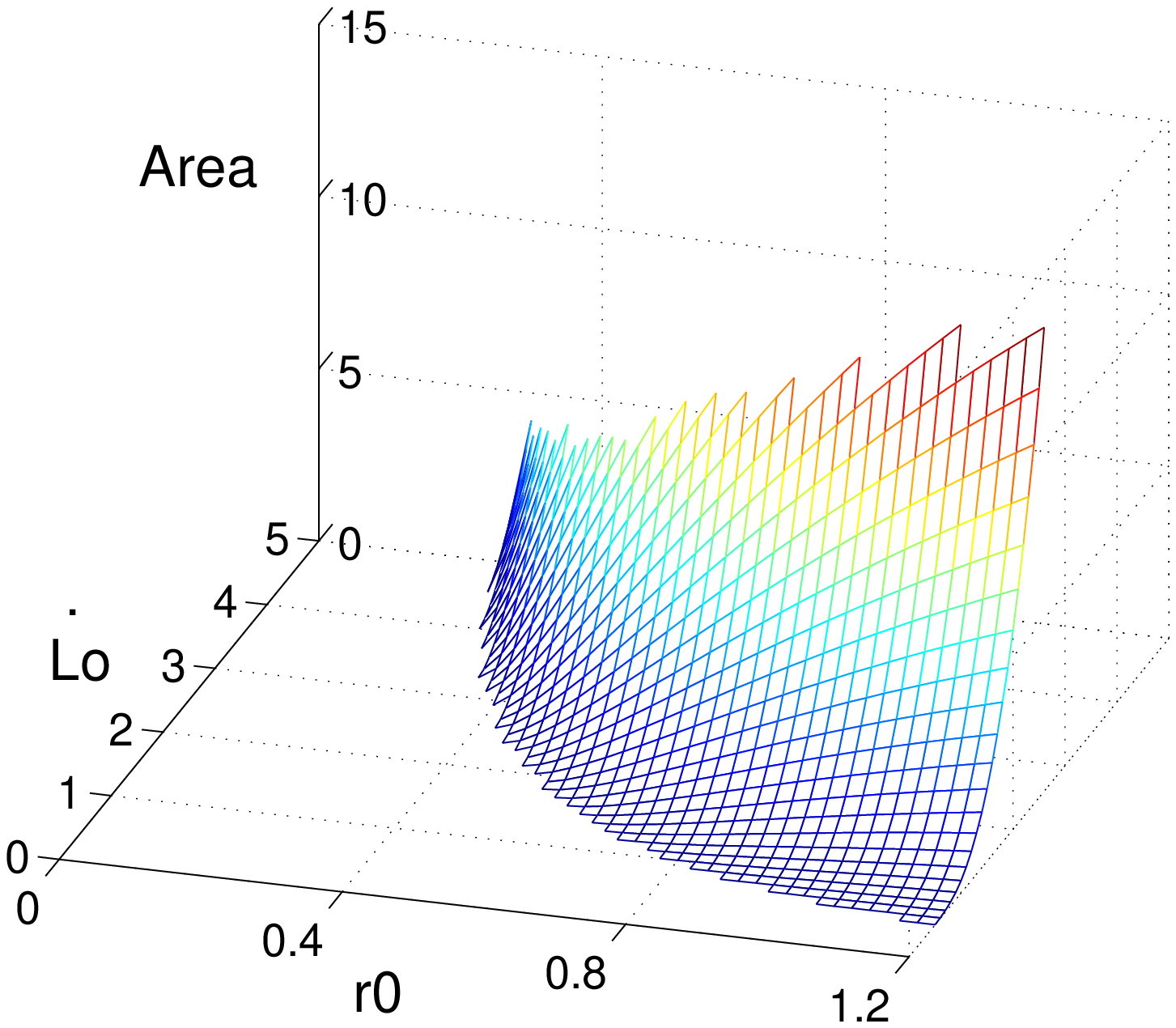}
\caption{The effects on the creation of a black hole and it's final area due to differing the initial $\dot{L_0}$ and r0.} 
\lab{fig:param}}

\section{Conclusions}

The possibility of transitions which change the topology of a Calabi Yau manifold, is intriguing. 
It suggests a method of connecting, seemingly discrete, moduli spaces of Calabi Yau manifolds by 
their common singularities. If the transitions can occur dynamically, then the topology of the 
compactified dimensions of string theory may change in time. Based on this expectation one can
study the low-energy dynamics coming from such a transition.

To study these topology changing processes we perform numerical simulations of an Eguchi-Hanson
spacetime with a collapsing two-cycle.
Our results highlight the importance of gravity during these events where cycles are collapsing,
showing that either horizons form during the process, or the cycle re-expands. In either case
one concludes that the gravitational effects prevent the naive collapse of the cycle.

We have presented evidenvce that, in the case where a horizon forms, the final state of the evolution
is a black hole, where the horizon is a squashed three-sphere \cite{Gibbons:1985ac,Dobiasch:1981vh}.
Such a black-hole has an interesting asymptotic structure, namely there is a compact circle at 
infinity, and this leads us to an unexpected mechanism for compactificaction. 
If, instead of picturing the Eguchi-Hanson
space as a portion of a compact internal space, we start with the full Eguchi-Hanson space, with its
four ``large'' dimensions, we see that the final state has a compact dimension and corresponds to the
Kaluza-Klein black-hole of \cite{Gibbons:1985ac}.

\acknowledgments We thank James Gray, Hari Kunduri and Harvey Reall for useful discussions. Both
NAB and PMS are supported by STFC.

\vskip 1cm
\appendix{\noindent\Large \bf Appendices}
\section{Constraints}
\lab{app:constraints}

The metric produced the following constraints, equations the variables must always conform to. These were imposed as 
initial conditions and later monitored to test the program's accuracy.

The Hamiltonian constraint:
\bea
0=&&-K_A\,(2\,K_B+K_C)-K_B\,(K_B+2\,K_C)+\frac{c^2}{b^4}-\frac{4}{b^2}
\lab{HamCon}\\
&&+\frac{1}{a^2}\left(-D_A\,(D_C+\frac{1}{r})+2\,D_B\,(D_C+\frac{1}{r})
 -2\,D_A\,D_B+3\,D_B^2+2\,D_B'+D_C'+\frac{2\,D_C}{r}+D_C^2\right)\nn
\eea

Also the momentum constraint:
\be
K_A\,(D_C+2\,D_B+\frac{1}{r})=2\,K_B'+K_C'+2\,K_B\,D_B+K_C\,(D_C+\frac{1}{r}).
\lab{MomCon}
\ee
where:
\be
a^2=\left(1+4\left(\frac{r}{l}\right)^2\right)\,e^{(2A)}\,\,\,\,\,\,\,\,b^2=4l^2\,\left(1+\left(\frac{r}{l}\right)^4\right)\,e^{(2B)}\,\,\,\,\,\,\,\,c^2=4\,r^2\,\left(1+\left(\frac{r}{l}\right)^2\right)\,e^{(2C)}
\lab{Parameters}
\ee

Apparently singular terms within these constraints did not 
produce any instabilities as they do not feed back into 
the equations used to evolve the system, they were only used for testing purposes.
\section{Equations of motion}
\lab{app:eom}

We found the equations of motion from the field equations using the ADM formalism \cite{Arnowitt:1962hi,Kelly:2001kj}.
Also we added multiples of the momentum and Hamiltonian constraints 
to remove as many potentially singular terms from our equations of motion.
This resulted in the following equations of motion.

\bea
\dot{A}&=&-K_A\nn\\
\dot{B}&=&-K_B\nn\\
\dot{C}&=&-K_C\nn\\
\nn\\
\dot{K_A}&=&\frac{1}{a^2}(D_B^2+2\,D_B\,(D_C+\frac{1}{r}))+\frac{c^2}{b^4}-\frac{4}{b^2}+K_A^2-K_B^2-2\,K_B\,K_C\nn\\
\dot{K_B}&=&\frac{1}{a^2}(D_B\,D_A-2\,D_B^2-D_B'-D_B\,(D_C+\frac{1}{r}))
-\frac{2\,c^2}{b^4}+\frac{4}{b^2}+2\,K_B^2+K_B\,K_C+K_B\,K_A\nn\\
\dot{K_C}&=&\frac{1}{a^2}(3\,D_B^2-2\,D_B\,D_A+2\,D_B')+\frac{3\,c^2}{b^4}-\frac{4}{b^2}-K_B^2+K_C^2-2\,K_B\,K_A\nn\\
\nn\\
\dot{D_A}&=&-K_A'\nn\\
\dot{D_B}&=&-K_B'\nn\\
\dot{D_C}&=&-K_C'
\lab{EQmo}
\eea
Where $a$, $b$ and $c$ are defined in $(\ref{Parameters})$.
The only potentially singular term remaining (which could have produced instabilities) is $D_B/r$, 
analytically this is regular as $D_B$ is odd. 
Numerically it was sufficiently stable to allow the program to run its course.

\end{document}